\documentclass[preprint,double-spaced,showpacs,amsmath,amssymb]{revtex4}
\usepackage{graphicx}
\usepackage{dcolumn}

\begin{document}

\newcommand{\beq}{\begin{equation}}
\newcommand{\eeq}{\end{equation}}
\newcommand{\bea}{\begin{eqnarray}}
\newcommand{\eea}{\end{eqnarray}}
\newcommand{\eps}{\varepsilon}
\newcommand{\lsim}{\stackrel{\scriptstyle <}{\phantom{}_{\sim}}}
\newcommand{\gsim}{\stackrel{\scriptstyle >}{\phantom{}_{\sim}}}
\newcommand{\bfg}{\boldsymbol}

\title{ Magnetic moments of long isotopic chains}

\author{ I.\,N. Borzov}
\email{i.borzov@gsi.de} \affiliation{ IPPE, Obninsk, Russia}

\author{ E.\,E. Saperstein}

\author{ S.\,V. Tolokonnikov}
\email{tolkn@mbslab.kiae.ru} \affiliation{Kurchatov Institute,
123182, Moscow, Russia}

\author{ G. Neyens}

\author{N. Severijns}
\affiliation{Instituut voor Kern- en Stralingsfysica, Katholieke
Universiteit Leuven,  B-3001 Leuven, Belgium }

\date{\today}

\begin{abstract}
Dipole magnetic moments of several long isotopic chains are
analyzed  within the self-consistent Finite Fermi System theory
based on the Generalized Energy Density Functional method with
exact account for the pairing and quasi-particle continuum. New
data for nuclei far from the $\beta$-stability valley are included
in the analysis. For a number of semi-magic isotopes of the tin
and lead chains  a good description of the data is obtained, with
accuracy of $0.1{-}0.2 \mu_{\rm N}$. A chain of non-magic isotopes
of  copper is also analyzed  in detail. It is found that the
systematic analysis of magnetic moments of this long chain yields
rich information on the evolution of the nuclear structure  of the
Cu isotopes. In particular, it may give a signal of deformation
for the ground state of some  nuclei in the chain.
\end{abstract}


\maketitle

\section{Introduction}
Dipole magnetic moments of atomic nuclei   are one of the
fundamental nuclear characteristics. Their interpretation played
an important role in the formulation of basic theoretical
approaches in nuclear physics,  such as the Shell Model (SM)
\cite{Bohr} and the Finite Fermi System (FFS) theory \cite{AB1}.
Modern RIB facilities provide an access to long chains of isotopes
including the radioactive ones in their ground and isomeric
states. Spectroscopy techniques using high intensity lasers allow
for precision measurements of  nuclear spins and magnetic moments.
As a result, the bulk of the data on nuclear magnetic moments
becomes very  extensive and comprehensive \cite{Stone} creating a
challenge to nuclear theory.

Recently, the self-consistent version of the FFS theory was
applied to describe magnetic moments of more than 100   odd-A
spherical heavy and intermediate atomic mass nuclei \cite{BST}.
This approach is based on the so-called Generalized Energy Density
Functional (EDF) method for nuclei with pairing correlations
\cite{STF,Fay}. It involves explicit consideration of the pion and
$\rho$-meson degrees of freedom and the method of finding the FFS
(RPA-like) response function with exact account for all continuum
states. The latter was developed in \cite{ShB,SapTF} for magic
nuclei and in \cite{Plat}  for the general case taking into
account pairing correlations. With the exception of a number of
cases, a high precision description with accuracy of $0.1{-}0.2
\mu_{\rm N}$ was obtained in \cite{BST}. In these calculations the
``one-quasiparticle'' approximation was used in which an odd
nucleus is considered as a system with a quasiparticle  added to
the even-even core in the state $\lambda_0$ at the Fermi surface.
In the FFS theory, the quasiparticle differs from the particle in
the SM in two respects. First, it possesses the effective local
quasiparticle charge $e_q[V_0]$ for the  external field $V_0$
under consideration. This charge is, as a rule, different  from
that of the bare particle. Second, the core is polarized with the
quasiparticle, via the so-called Landau--Migdal interaction
amplitude ${\cal F}$, playing the role of the effective
interaction in the FFS theory. In this approach, both effects are
taken into account in the equation for the effective field $V$ in
terms of the bare field $V_0$, $e_q[V_0]$ and ${\cal F}$, the
magnetic moment value being given by the single-particle matrix
element $ \mu_{\lambda_0 }= (V[\hat{\bfg\mu}])_{\lambda_0
\lambda_0}$. It should be noted that in \cite{BST}, in addition to
the two standard FFS theory local charges with respect to the
magnetic field, i.e. the spin $\zeta_s$ and orbit $\zeta_l$ local
charges, a new, tensor local charge $\zeta_t$ was added. It is
defined in such a way that the corresponding contribution to the
magnetic moment operator is $\delta
\mu_t=\zeta_t\sqrt{4\pi}\,[{\bf Y}_2({\bf
n})\circledast\hat{\boldsymbol\sigma}]^1_\mu\,\hat{\tau}_3$, where
${\bf n}={\bf r}/r$, and $\hat{\tau}_3$ is the third isospin Pauli
matrix. It turned out that introducing such local charge with
$\zeta_t=0.2$ makes the overall agreement of the theory with
experiment better, especially for nuclei with the odd nucleon in
the state $\lambda_0=p_{1/2}$. Other FFS theory parameters
entering the calculation scheme in the case of the magnetic
symmetry were taken from \cite{PF,BorTF}.

In this paper, within the same approach as in \cite{BST} , we
concentrate on the analysis of magnetic moments of long isotopic
chains. Recently  a lot of experimental data have been obtained
for Cu isotopes \cite{Minam,Isolde_Cu,Cu_1}  and Pb-isotopes
\cite{Pb} . We have analyzed the cases for which a simple approach
\cite{BST} based on the assumption of sphericity and on the
one-quasiparticle approximation is not compatible with data. It
was found that in several cases the contribution of more
complicated configurations should be taken into account. Further,
for nuclei at the extremes of the isotopic chains the deformation
was found to play a role as well.

\section{Semi-magic nuclei}
Let us begin from the analysis of the global behavior of magnetic
moments for the two chains of semi-magic isotopes of tin ($Z=50$)
and lead ($Z=82$). Ground state energies and radii of both
isotopic chains were analyzed  in detail by S.A. Fayans et al.
\cite{Fay} within the Generalized EDF method \cite{STF} which is a
generalization  to superfluid systems of the well-known EDF method
by Kohn and Sham \cite{KS} developed originally for normal Fermi
systems.  The Sn and Pb chains were analyzed assuming a spherical
shape for the ground state. The very good description of masses
and radii, including the odd-even effects, confirmed this
assumption. This is in agreement with the systematic analysis of
nuclear deformations carried out in \cite{HF} within the Advanced
Thomas-Fermi method with the Shell-Model correction  of V.M.
Strutinsky. It should be noted that we use the term ``spherical''
for odd nuclei neglecting a small ($\propto 1/A$) deformation
effect of the odd nucleon. In fact, this is  taken into account in
the FFS theory equations for the effective field $V$ by
calculating the core polarization induced  by this particle.

We start with the Sn chain containing 12 odd isotopes, from
 $^{109}$Sn to $^{131}$Sn, with neutron number varying from $N_{\rm min}=59$
 to $N_{\rm max}=81$. Note that we consider only nuclei with
 known experimental values of the magnetic moment in the ground or
 excited state. Among these nuclei, only $^{115-119}$Sn are stable, i.e.
 the $\beta$-stability valley for odd Sn isotopes has boundaries
 at  $N^\beta_{\rm min}=65$ and $N^\beta_{\rm max}=69$. Thus, the maximum
 neutron excess is $\delta N_+ = N_{\rm max}-N^\beta_{\rm max}=12$
 and the deficiency, $\delta N_- = N^\beta_{\rm min} - N_{\rm min}=6$. It is
 instructive to consider also the relative maximum neutron
 excess, $\eps_+=\delta N_+/N^\beta_{\rm max}=17.4\%$, and deficiency,
$\eps_-=\delta N_-/N^\beta_{\rm min}=9.2\%$.

\begin{figure}[]
\centerline{\includegraphics [height=120mm]{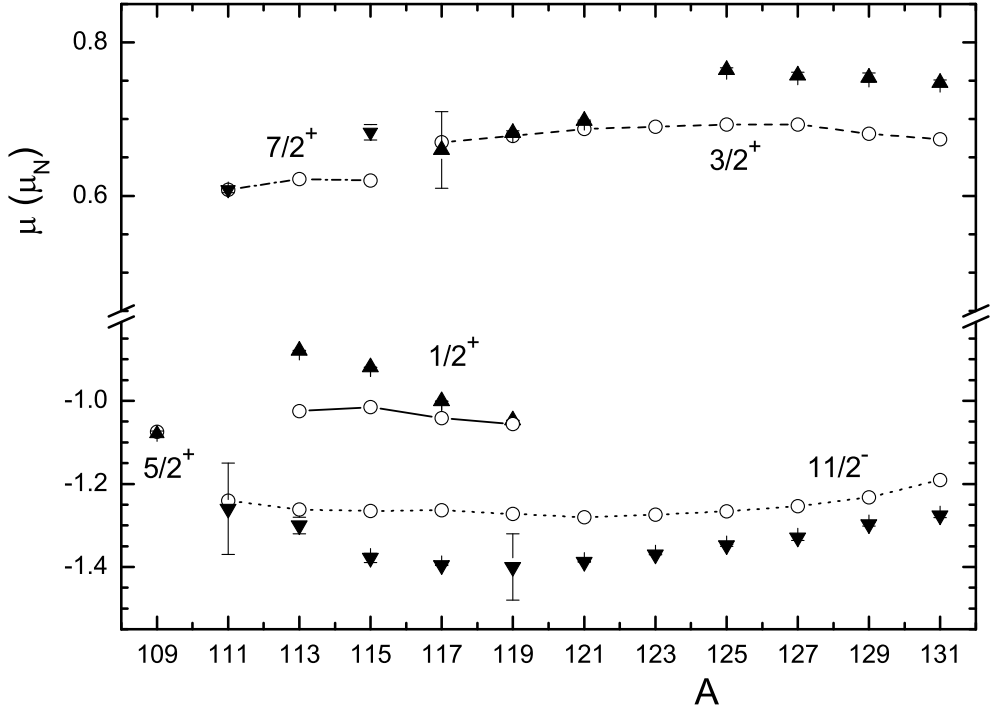}}
\caption{Magnetic moments of Sn isotopes. The  experimental data
are shown with filled triangles (up and down) \cite{Stone}. Open
circles show the theoretical predictions. The quantum numbers of
the  single-particle configurations   assumed in our calculations
are also indicated.}\label{mu_Sn}
\end{figure}

A comparison of 25 theoretical values of magnetic moments with the
corresponding experimental data  taken from \cite{Stone} is given
in Table I and Fig. \ref{mu_Sn}.
 For convenience, we present also the difference
$\delta \mu = \mu_{\rm th} - \mu_{\rm exp}$ which characterizes
directly the   global accuracy of the theory. As we see, this
accuracy is sufficiently high: as a rule, we obtained $|\delta
\mu| < 0.1 \mu_{\rm N}$, and only in four cases $ 0.1 \mu_{\rm N}
< |\delta \mu| < 0.15 \mu_{\rm N}$.  Except for two out of 25
cases (i.e. $^{113}$Sn and $^{115}$Sn) all experimental magnetic
moments are reproduced within 10\%.

\begin{table}
\caption{Magnetic moments (in $\mu_{\rm N}$) of Sn isotopes.
         The experimental data are from \cite{Stone}.
         The symbol * indicates  excited states.}

\begin{ruledtabular}
\begin{tabular}{cccccc}
 nucleus &  neutron s.p. state $\lambda_0^n$ &
$\mu_{\rm exp}$  & $\mu_{\rm Sch}$& $\mu_{\rm th}$
&$\delta \mu$\\
 \hline
$^{109}{\rm Sn}$ &$d_{5/2}$ & -1.079(6)& -1.913     & -1.074 &  0.005(6) \\
\hline
$^{111}{\rm Sn}$ &$g_{7/2}$ & 0.608(4)& 1.488       & 0.608  &  0.000(4) \\
$^{115}{\rm Sn}^*$ &$g_{7/2}$ & 0.683(10)& 1.488       & 0.620  &  -0.06(1) \\
\hline
$^{113}{\rm Sn}$ &$s_{1/2}$ & -0.8791(6)& -1.913      & -1.025 & -0.146 \\
$^{115}{\rm Sn}$ &$s_{1/2}$ & -0.91883(7)& -1.913     & -1.015 & -0.096 \\
$^{117}{\rm Sn}$ &$s_{1/2}$ & -1.00104(7)& -1.913     & -1.042 & -0.041 \\
$^{119}{\rm Sn}$ &$s_{1/2}$ & -1.04728(7)& -1.913     & -1.056 & -0.009 \\
\hline
$^{111}{\rm Sn}^*$ &$h_{11/2}$ & -1.26(11)& -1.913  & -1.240 &  0.02(11) \\
$^{113}{\rm Sn}^*$ &$h_{11/2}$ & -1.30(2)& -1.913  & -1.262 &  0.04(2) \\
$^{115}{\rm Sn}^*$ &$h_{11/2}$ & -1.378(11)& -1.913  & -1.265 &  0.11(1) \\
$^{117}{\rm Sn}^*$ &$h_{11/2}$ & -1.3955(10)& -1.913  & -1.263 &  0.132 \\
$^{119}{\rm Sn}^*$ &$h_{11/2}$ & -1.40(8)& -1.913  & -1.272 &  0.13(8) \\
$^{121}{\rm Sn}^*$ &$h_{11/2}$ & -1.3877(9)& -1.913   & -1.280 &  0.108 \\
$^{123}{\rm Sn}$ &$h_{11/2}$ & -1.3700(9)& -1.913     & -1.274 &  0.096 \\
$^{125}{\rm Sn}$ &$h_{11/2}$ & -1.348(2)& -1.913      & -1.266 &  0.082(2) \\
$^{127}{\rm Sn}$ &$h_{11/2}$ & -1.329(7)& -1.913      & -1.254 &  0.075(7) \\
$^{129}{\rm Sn}^*$ &$h_{11/2}$ & -1.297(5)& -1.913      & -1.232 &  0.065(5) \\
$^{131}{\rm Sn}^*$ &$h_{11/2}$ & -1.276(5)& -1.913      & -1.191 &  0.085(5) \\
\hline
$^{117}{\rm Sn}^*$ &$d_{3/2}$ & 0.66(5)& 1.148       & 0.670  & 0.01(5)   \\
$^{119}{\rm Sn}^*$ &$d_{3/2}$ & 0.682(3)& 1.148       & 0.678  & -0.004(3)   \\
$^{121}{\rm Sn}$ &$d_{3/2}$ & 0.6978(10)& 1.148       & 0.687  & -0.011(1)   \\
$^{125}{\rm Sn}^*$ &$d_{3/2}$ & 0.764(3)& 1.148       & 0.693  & -0.071(3)   \\
$^{127}{\rm Sn}^*$ &$d_{3/2}$ & 0.757(4)& 1.148       & 0.693  & -0.064(4)   \\
$^{129}{\rm Sn}$ &$d_{3/2}$ & 0.754(6)& 1.148         &  0.681 & -0.073(6) \\
$^{131}{\rm Sn}$ &$d_{3/2}$ & 0.747(4)& 1.148         &  0.674 & -0.073(4) \\
\end{tabular}
\end{ruledtabular}
\end{table}

\begin{figure}[]
\centerline{\includegraphics [height=120mm]{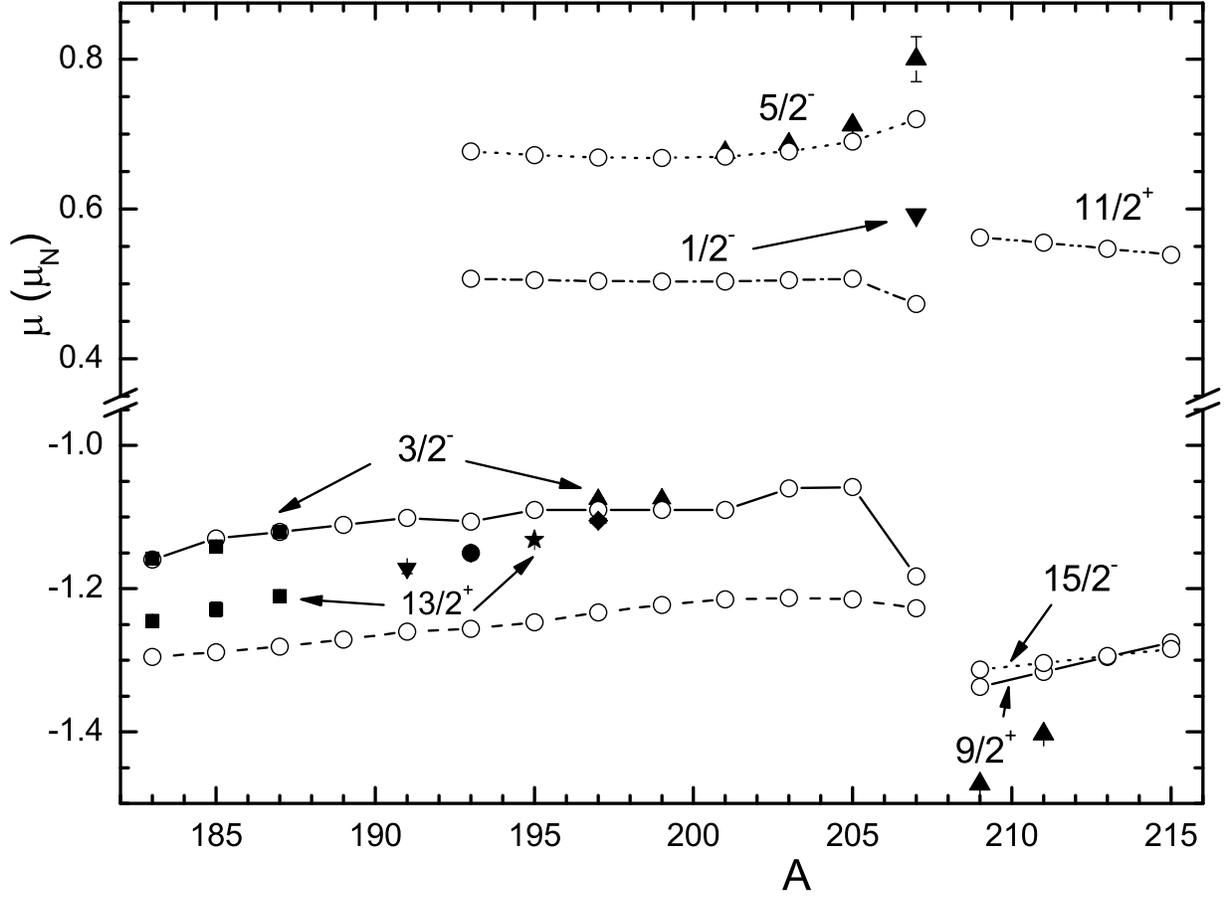}} \caption{
Magnetic moments of Pb isotopes. The experimental data are shown
with filled triangles (up and down) \cite{Stone}, squares
\cite{Pb}, filled circles \cite{Dutta}, a star \cite{Dinger} and a
rhombus \cite{Anselment}. Open circles show the theoretical
predictions. The quantum numbers of the single-particle
configurations assumed in the calculations are also indicated.
}\label{mu_Pb}
\end{figure}

Now turn to the Pb chain containing 15 odd isotopes, from
$^{183}$Pb to $^{211}$Pb ($N_{\rm min}=101$, $N_{\rm max}=129$).
In this case, there is only one odd $\beta$-stable isotope
$^{207}$Pb, hence $N^\beta_{\rm min}=N^\beta_{\rm max}=101$. The
maximum neutron excess is equal to  $\delta N_+ = 4$, $\eps_+ =
4\%$ and the deficiency, $\delta N_- = 25$, $\eps_- = 24.8\%$. The
theoretical predictions for 19 experimental values of magnetic
moments  \cite{Stone,Pb,Dutta,Dinger,Anselment} are presented in
Table II and Fig.\ref{mu_Pb}. Again the accuracy of the theory is
rather good: $|\delta \mu| < 0.1 \mu_{\rm N}$ in 15 cases and $
0.1 \mu_{\rm N} < |\delta \mu| < 0.15 \mu_{\rm N}$ in four  
cases.  Except for 2 out of 13 cases (i.e. $^{197}$Pb$^*$ and
$^{207}$Pb) all experimental moments are reproduced within 10\%.

\begin{table}
\caption{Magnetic moments (in $\mu_{\rm N}$) of Pb isotopes. The
experimental data are from \cite{Stone}, \cite{Dutta} $(^{a)})$,
\cite{Dinger} $(^{b)})$ and \cite{Anselment} $(^{c)})$. Those in
preparation are given by Ref. \cite{Pb}. The symbol
* indicates  excited states.}

\begin{ruledtabular}
\begin{tabular}{cccccc}
 nucleus &  neutron s.p. state $\lambda_0^n$ &
$\mu_{\rm exp}$  & $\mu_{\rm Sch}$& $\mu_{\rm th}$
&$\delta \mu$\\
 \hline
 $^{183}{\rm Pb}$ &$p_{3/2}$ &\cite{Pb} & -1.913       & -1.139 & -  \\
$^{185}{\rm Pb}$ &$p_{3/2}$ & \cite{Pb}& -1.913       & -1.130 & -  \\
$^{187}{\rm Pb}$ &$p_{3/2}$ & \cite{Pb}& -1.913       & -1.121 & -  \\
$^{197}{\rm Pb}$ &$p_{3/2}$ & -1.075(2)& -1.913     & -1.090 & -0.015(2) \\
$^{199}{\rm Pb}$ &$p_{3/2}$ & -1.0742(12)& -1.913     & -1.090 & -0.016(1) \\
\hline
$^{183}{\rm Pb}^*$ &$i_{13/2}$ & \cite{Pb}& -1.913      & -1.295 & - \\
$^{185}{\rm Pb}^*$ &$i_{13/2}$ &\cite{Pb}& -1.913     & -1.289 &  - \\
$^{187}{\rm Pb}^*$ &$i_{13/2}$ & \cite{Pb}& -1.913      & -1.281 & - \\
$^{191}{\rm Pb}$ &$i_{13/2}$ & -1.172(7)& -1.913    & -1.260 & -0.088(7) \\
$^{193}{\rm Pb^{a)}}^*$ &$i_{13/2}$ & -1.150(7)& -1.913    & -1.256 & -0.106(7) \\
$^{195}{\rm Pb^{b)}}^*$ &$i_{13/2}$ & -1.1318(13)& -1.913  & -1.247 & -0.115(1) \\
$^{197}{\rm Pb^{c)}}^*$ &$i_{13/2}$ & -1.105(3)& -1.913    & -1.233 & -0.128(3) \\
 \hline
$^{201}{\rm Pb}$ &$f_{5/2}$ & 0.6753(5)& 1.366        & 0.670  & -0.005 \\
$^{203}{\rm Pb}$ &$f_{5/2}$ & 0.6864(5)& 1.366        & 0.677  & -0.009 \\
$^{205}{\rm Pb}$ &$f_{5/2}$ & 0.7117(4)& 1.366        & 0.690  & -0.022 \\
$^{207}{\rm Pb}^*$ &$f_{5/2}$ & 0.80(3)      & 1.366  & 0.720  & -0.08(3)\\
\hline
$^{207}{\rm Pb}$ &$p_{1/2}$   & 0.592585(9)   & 0.638  & 0.473  & -0.120\\
\hline
$^{209}{\rm Pb}$ &$g_{9/2}$   & -1.4735(16)   & -1.913 & -1.337 &  0.137(2)\\
$^{211}{\rm Pb}$ &$g_{9/2}$ & -1.4037(8)& -1.913      & -1.316 &  0.088(1) \\

\end{tabular}
\end{ruledtabular}
\end{table}

Thus, for the chains of semi-magic  Sn and Pb isotopes containing
nuclides that are sufficiently far from the $\beta$-stability
valley (the maximum value of the relative distance is $\eps \simeq
25\%$), the   two main assumptions of the calculation procedure,  
i.e. the spherical symmetry and one-quasiparticle approximation,
work sufficiently well. The agreement between the theory and
experimental data confirms also the universal character of the
spin dependent parameters of the extended FFS theory which were
found primarily for $\beta$-stable nuclei. Similar conclusions
were obtained also in   systematic studies of the $\beta$-decay
properties of very neutron-rich nuclei \cite{Bo}.

\begin{figure}[]
\centerline{\includegraphics [height=120mm]{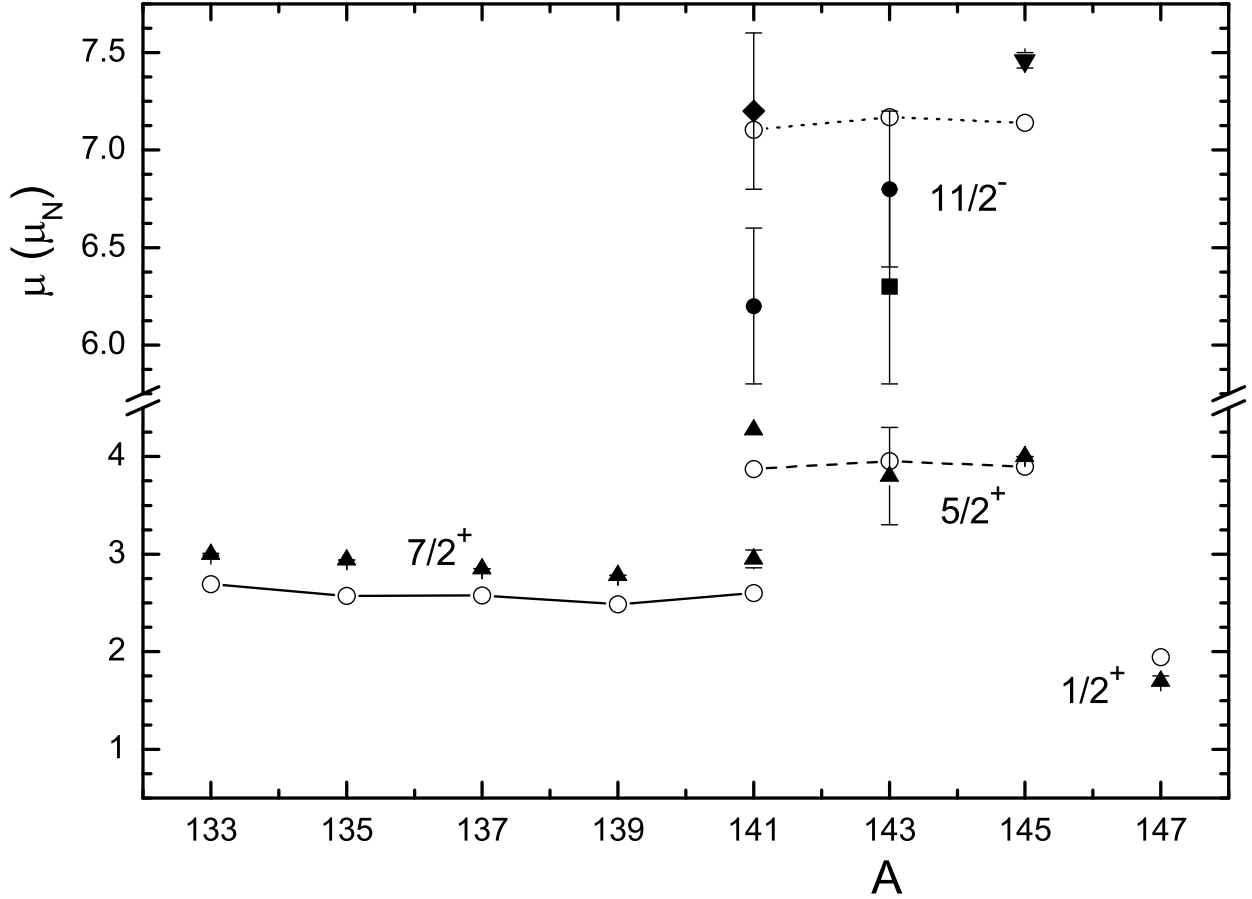}} \caption{
Magnetic moments of  N=82 isotones.The  experimental data are
shown with filled triangles (up and down) \cite{Stone}, filled
circles \cite{Go12}, a rhombus \cite{Ej01} and a square
\cite{Pr02}. Open circles show the theoretical predictions. The
quantum numbers of the assumed single-particle configurations are
also indicated.}\label{mu_n82}
\end{figure}

For completeness, we present also the magnetic moments of isotones
with $N=82$, $Z_{\rm min}=51$ and $Z_{\rm max}=65$. They are
collected in Table III and Fig.\ref{mu_n82}. In this case, only  
the nuclei $^{139}$La and $^{141}$Pr are $\beta$-stable, i.e. the
$\beta$-stability valley for odd $N=82$ isotones has limits
$Z_{\rm min}=57$ and $Z_{\rm max}=59$. Using the notation similar
to the one introduced above, one finds $\delta Z_- = \delta Z_+ =
6$, and $\eps'_- \simeq \eps'_+ \simeq 10\%$.     It can be seen
from Fig.3 and Table III that in terms of absolute deviation
$\delta \mu$ the accuracy of the theoretical predictions is
somewhat worse than for the isotopic chains analyzed above.
Nevertheless, the theoretical results differ at most 15\% from the
experimental values. For some of these isotones which are rather
close to the lanthanide deformed region the main assumptions of
these calculations are not necessarily valid.    

\begin{table}
\caption{Magnetic moments (in $\mu_{\rm N}$) of N=82 isotones. The
experimental data are from \cite{Stone}, \cite{Go12} $(^{a)})$,
\cite{Ej01} $(^{b)})$ and \cite{Pr02} $(^{c)})$. The symbol
* indicates an excited state.}

\begin{ruledtabular}
\begin{tabular}{cccccc}
 nucleus &  proton s.p. state $\lambda_0^p$ &
$\mu_{\rm exp}$  & $\mu_{\rm Sch}$& $\mu_{\rm th}$
&$\delta \mu$\\
 \hline

$^{133}{\rm Sb}$ &$g_{7/2}$ & 3.00(1)& 1.817          &  2.693 & -0.31(1) \\
$^{135}{\rm I}$ &$g_{7/2}$ & 2.940(2)& 1.817          &  2.570 & -0.370(2) \\
$^{137}{\rm Cs}$ &$g_{7/2}$ & 2.8513(7)& 1.817          &  2.577 & -0.274 \\
$^{139}{\rm La}$ &$g_{7/2}$ & 2.7830455(9)& 1.817     &  2.488 & -0.295 \\
$^{141}{\rm Pr}^*$ &$g_{7/2}$ & 2.95(9)& 1.817          &  2.603 & -0.35(9) \\
\hline
$^{141}{\rm Pr}$ &$d_{5/2}$ & 4.2754(5)& 4.793        &  3.871 & -0.404 \\
$^{143}{\rm Pm}$ &$d_{5/2}$ & 3.8(5)& 4.793           &  3.954 &  0.1(5) \\
$^{145}{\rm Eu}$ &$d_{5/2}$ & 3.999(3)& 4.793         &  3.894 & -0.105(3) \\
\hline
$^{141}{\rm Pr^{a)}}^*$ &$h_{11/2}$ & 6.2(4)& 7.793 &  7.103 & 0.9(4) \\
$^{141}{\rm Pr^{b)}}^*$ &$h_{11/2}$ & 7.2(4)& 7.793 &  7.103 & -0.1(4) \\
$^{143}{\rm Pm^{a)}}^*$ &$h_{11/2}$ & 6.8(4)& 7.793        &  7.168 &  0.4(4) \\
$^{143}{\rm Pm^{c)}}^*$ &$h_{11/2}$ & 6.3(5)& 7.793        &  7.168 &  0.9(5) \\
$^{145}{\rm Eu}^*$ &$h_{11/2}$ & 7.46(4)& 7.793       &  7.141 & -0.32(4) \\
\hline
$^{147}{\rm Tb}$ &$s_{1/2}$ & 1.70(5)& 2.793          &  1.942 &  0.24(5)  \\
\end{tabular}
\end{ruledtabular}
\end{table}

\section{The copper chain}

Now we consider an example  of a chain of non-magic isotopes, the
copper one. It contains 11 odd isotopes, from $^{57}$Cu to
$^{77}$Cu ($N_{\rm min}=28$, $N_{\rm max}=48$). In this case, two
isotopes are $\beta$-stable, $^{63,65}$Cu, i.e. we have
$N^\beta_{\rm min}= $34, $N^\beta_{\rm max}=36$. The corresponding
characteristics of the neutron excess and deficiency are equal to
$\delta N_+ = 12$, $\eps_+ = 33.3\%$, $\delta N_- = 6$, $\eps_- =
17.6\%$. Thus, although the absolute value of the distance from
the $\beta$-stability valley is less  than that for the semi-magic
chains considered in the previous section, the relative  neutron
excess is now maximal. Seven of these isotopes, $^{57-69}$Cu,
belong to the $2p-1f$ shell, while in the isotopes $^{71-77}$Cu,
the next higher $1g_{9/2}$ shell is filling. A systematic analysis
of the properties of nuclei of the $2p-1f$ shell within the
many-particle SM was carried out before \cite{Brown}.
Unfortunately, this paper contains only three isotopes of the Cu
chain, i.e. with $A=61,63,65$.

We now analyzed the entire Cu isotopic chain using different
assumptions.A comparison of the calculated values with
experimental data \cite{Stone,Isolde_Cu,Cu_1,UliKos} is given in
Table~\ref{Cu} and displayed in Fig.~\ref{mu_Cu}.

\subsection{Spherical calculations for copper chain}

\begin{figure}[]
\centerline{\includegraphics [height=120mm]{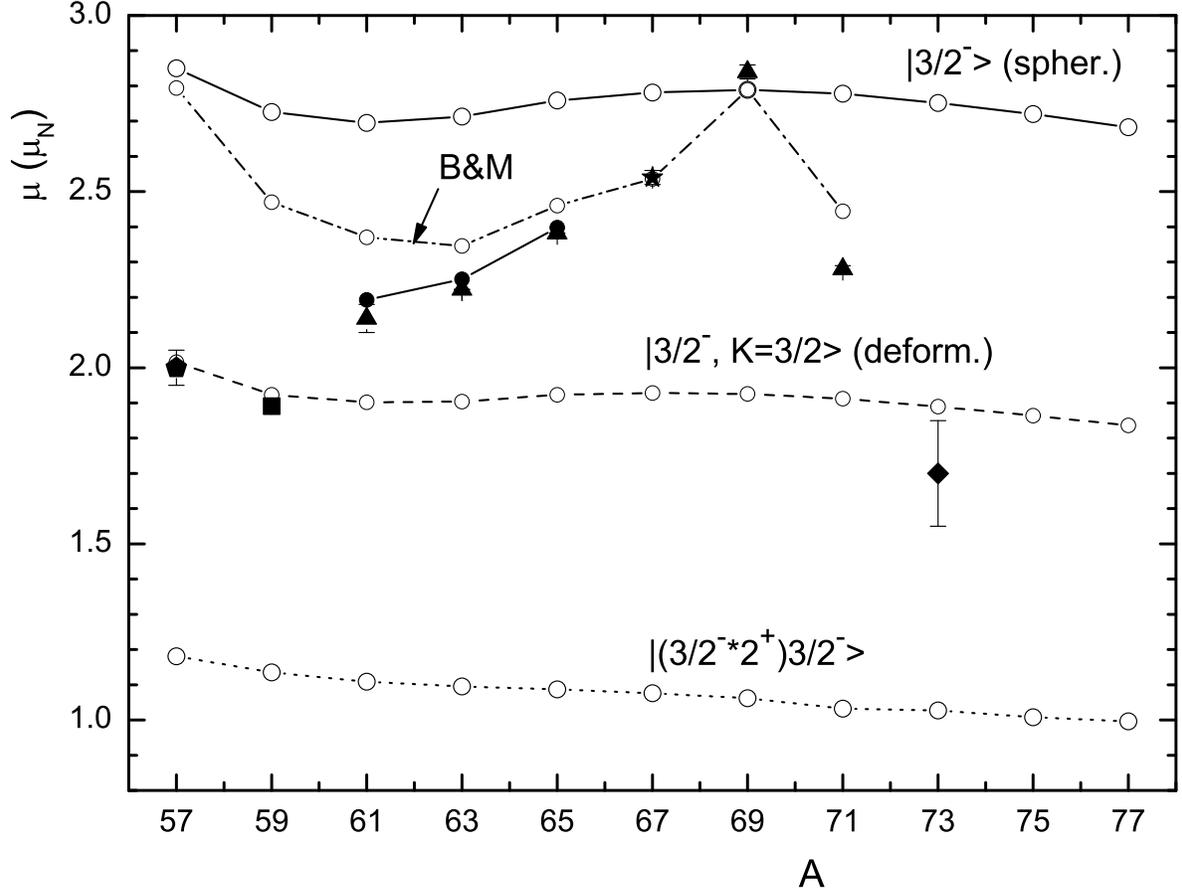}} \caption{
Magnetic moments of Cu isotopes. The  experimental data are shown
with filled triangles \cite{Stone}, a filled pentagon
\cite{Minam}, a filled square \cite{Isolde_Cu}, a filled star
\cite{RiStone}, or a filled rhombus \cite{UliKos}.  Theoretical
values from \cite{Brown} ($I=3/2$) are shown with filled circles.
The assumed configurations are indicated for spherical, deformed
and phonon-particle configurations, correspondingly (see details
in text).}\label{mu_Cu}
\end{figure}

First, one can see that the spherical calculations performed
within the same one-quasiparticle scheme as above (column 6 of
Table~\ref{Cu}  and open circles labelled $|3/2^-\rangle$(spher)
in Fig.\ref{mu_Cu}) may have some relevance only in the middle of
the chain, i.e. near A=69 corresponding to the magic number N=40.
For $^{61-65}$Cu, the full fp-space predictions within the
many-particle SM \cite{Brown}  (column 5) are in perfect agreement
with the data. The fact that the simple FFS theory calculations
deviate from the experimental data indicates that many-particle
admixtures should be taken into account in this region. The
magicity of N=40 and related characteristics of the low-lying
quadrupole excitations were discussed in \cite{Kaneko}. Our
spherical calculation gives some grounds to assume an onset of
deformation in $^{57,59,73}$Cu, this possibility will be discussed
in the section D.  For $^{75}$Cu and $^{77}$Cu isotopes, the
experimental ground state $I^{\pi}$ values and magnetic moments
can in principle be measured.

\subsection{Particle-phonon coupling calculations}

Among the relevant mechanisms of renormalization of the magnetic
moments in spherical nuclei one should consider the correction
induced by the virtual excitation of the low-laying states (the
``phonons''). A consistent calculation of such contributions with
the many-body theory methods is very complicated. Up to now, it
was accomplished into practice only for magic nuclei where pairing
is absent \cite{KhS}. As an approximate version, we  use the
simple particle-phonon coupling model \cite{Bohr2} developed for
the magnetic moment problem by I. Hammamoto \cite{Hammam}. In this
model, the unperturbed ground state $|jm>$ of the odd nucleus
under consideration is mixed with the  unperturbed particle-phonon
one, \beq |(jL)jm>=\sum\limits_M C^{jm}_{jm-M\,LM}|jm-M>\,|LM>\,,
\eeq where $L$ is the phonon multipolarity. The magnetic moment of
such ``bare'' state is equal to \beq
\mu_{p+ph}=<(jL)jj|\,\hat{\mu}^z_p+\hat{\mu}^z_{ph}\,|(jL)jj>\,,
\eeq where $\hat{\mu}^z_{ph}$ stands for the 3-rd component of the
phonon magnetic moment operator $
\hat{\boldsymbol\mu}_{ph}=g_R\hat{\bf L}$, with $g_R$ being the
phonon gyromagnetic ratio. In the collective model used in
\cite{Hammam}, one has $g_R= Z/A$. A simple calculation yields
\beq \mu_{p+ph}=\left[1-\frac{L(L+1)}{2j(j+1)}\right]\mu_p
+\frac{L(L+1)}{2(j+1)}g_R\,. \eeq As we will see, for the Cu
nuclei the admixture of the particle-phonon state is often large,
and therefore the perturbation theory used in \cite{Hammam} is not
valid in this case. Instead, one could solve the two-level problem
we deal   with here  exactly. The dynamical admixture of the two
states under consideration is  \beq
\widetilde{|jm>}=c_1|jm>+c_2|(jL)jm>\,. \label{c12}\eeq The
corresponding magnetic moment is \beq
\tilde\mu_p=(1-w)\mu_p+w\mu_{p+ph}\,,\quad w=|c_2|^2=1-|c_1|^2,
\eeq or \beq \tilde\mu_p=\mu_p-w\left[\frac{L(L+1)}{2j(j+1)}\mu_p-
\frac{L(L+1)}{2(j+1)}g_R\right]\,.\label{mupph} \eeq For the Cu
isotopes under consideration the $2^+_1$-state contribution
dominates, such that for $\lambda_0=p_{3/2}$ we have $j=3/2\,,
L=2$, and \beq \tilde\mu_p=\mu_p-w\left(\frac45 \mu_p -\frac 65
g_R\right)\,.\label{mu2+} \eeq

To indicate the  maximal impact of the particle-phonon coupling we
display in Fig.~\ref{mu_Cu} the results for the maximum value of
the phonon mixing amplitude $w=1$ (i.e. the open circles labelled
as $|(3/2^{-}*2^+)3/2^-\rangle$). Next, for each nucleus we
calculated the value of $w$ which brings the magnetic moment into
agreement with experiment. These values are given in the 7-th
column of Table~\ref{Cu}.  It shows that near $A=69$ the required
phonon mixing amplitude is reasonably small (i.e. $w$ = 15--30\%).
On the wings of the chain {\bf $w$ $\geq$ 50\%}. In such a
situation, the model used is evidently irrelevant. At the left
wing of the chain, as well as at the right wing in the case of the
$I^{\pi}=3/2^-$ configuration, this can be interpreted as an
indication for the appearance of a stable ground state
deformation.

\subsection{A simple estimate of the phonon admixture
within the Bohr-Mottelson model}

A realistic estimate of the phonon admixture can also be obtained
within the collective Bohr-Mottelson (BM) model \cite{Bohr2} in
terms of the excitation energy $\omega_L$ and {\bf transition}
probability $B(EL)$ of the collective state under consideration.
Within this model, the particle-phonon interaction matrix element
reads:
 \beq
 h(j_1,(j_2L)j_1)=(-1)^L\left(\frac{2L+1}{4\pi}\right)^{1/2}
 C^{j_2\frac12}_{j_1\frac12\,L0}\left(\frac{\omega_L}{2C_L}\right)^{1/2}<j_2|k_L(r)|j_1>\,,
\label{h12}\eeq where $C_L$ is the stiffness coefficient of the
$L$-th vibration related to the vibration amplitude $\beta_L$  via
\beq \beta_L= \sqrt{2L+1}\left(\frac{\omega_L}{2C_L}\right)^{1/2},
\label{beta}\eeq and $k_L(r)$ stands for the radial form factor
which for  surface vibrations  has the form: \beq
<j_2|k_L(r)|j_1>\approx\left<j_2\left|R_0\frac{\partial
U}{\partial r}\right|j_1\right>\,. \label{form}\eeq For the states
in the vicinity of the Fermi level of the Cu isotopes these matrix
elements are approximately equal to 50 MeV. Inserting now in
Eq.~(\ref{h12}) $L$=2, $j_1$=$j_2$ =3/2, and using Eqs.~
(\ref{beta}) and (\ref{form}), one obtains a simple formula for
the Cu isotopes:
 \beq <h>\approx
\frac\beta{\sqrt{20\pi}}\,50\,[\mbox{MeV}]=6.31\beta
[\mbox{MeV}]\,.\eeq

Using the exact solution for the two-level problem, one finds for
the mixing probability \beq w_{\rm BM}
=\frac{|<h>|^2}{\left(\frac{\omega}2+\sqrt{\frac{\omega^2}4+
|<h>|^2}\right)^2+|<h>|^2}\,.\label{wrh} \eeq

We calculated values of $w_{\rm BM}$ (column 8 in Table~\ref{Cu})
for the $^{57-71}$Cu isotopes for which the experimental values of
$\omega_2$ and $\beta_2$ in the even-even Ni core are known (see
Table V). The last positions of the columns 8 and 9 are empty as
the corresponding data on the $2_1^+$-excitation in the $^{72}$Ni
nucleus are absent. To avoid double-counting of the effect, one
should take into account that the $2^+$-phonon admixture is
partially taken into account in the FFS theory equations via the
local quasiparticle charge $e_q$. On the basis of the analysis of
the magnetic moment of $^{69}$Cu, the odd neighbor of the magic
nucleus $^{68}$Ni, we find that the value of $\mu_p$ in
(\ref{mupph}) should be multiplied with the factor 1.05 (see
\cite{BST}). Taking this into account and using the values for
$w_{BM}$, Eq.~\ref{mupph} then yields the magnetic moments values
that are marked in Fig.~\ref{mu_Cu} with the label B\&M. As one
can see, the BM model helps to explain deviations from the
one-quasiparticle approximation result for nuclei in the vicinity
of $^{69}$Cu, the corresponding even-even core being close to
$^{68}$Ni. Indeed, in this case the phonon admixture coefficients
$w_{\rm BM}$ predicted by the BM model (column 8 of
Table~\ref{Cu}) are quite close to the ones ``required'' to fit to
data (the 7-th column of this table). This indicates the relevance
of the particle-phonon coupling mechanism in this region of the Cu
chain. On the other hand, for the $^{57,59}$Cu isotopes the values
of $w_{\rm BM}$ are much smaller than $w$, the latter being
unrealistically  large for these two isotopes. This indicates that
the particle-phonon mechanism is irrelevant in the vicinity of the
magic core $^{56}$Ni.

 Finally, it is of interest to mention that the calculations for
the isotopes with $A=61-65$ are in qualitative agreement with
those within the many-particle SM \cite{Brown}, indicating that
the 2$^+_1$-state admixture dominates among the many-particle
configurations.

\begin{table}
\caption{Magnetic moments (in $\mu_{\rm N}$) of  Cu isotopes. The
experimental data are from \cite{Stone}, \cite{Minam} $(^{a)})$,
\cite{Isolde_Cu} $(^{b)})$, \cite{RiStone} $(^{c)})$. \label{Cu} }

\begin{ruledtabular}
\begin{tabular}{ccccccccc}
 Nucleus & proton $\lambda_0$ & $\mu_{\rm exp}$ & $\mu_{\rm Sch}$ &
$\mu_{\rm th}$\cite{Brown}  & $\mu_{\rm th}$ & $w$ & $w_{\rm BM}$
&$\mu_{\rm th}$+$\delta \mu_{\rm BM}$\\

 \hline
$^{57}{\rm Cu^{a)ÿ}}$ &$p_{3/2}$ & 2.00(5) & 3.793 &   ---   &  2.850 &0.51&0.11 &  2.794\\
$^{59}{\rm Cu^{b)ÿ}}$ &$p_{3/2}$ & 1.891(9) & 3.793 &   ---  &  2.726 &0.53&0.23 &  2.470\\
$^{61}{\rm Cu}$ &$p_{3/2}$ & 2.14(4) & 3.793 & 2.193   &  2.695 &0.35&0.27 &  2.370\\
$^{63}{\rm Cu}$ &$p_{3/2}$ & 2.2273& 3.793 &2.251&  2.713 &0.30&0.29 &  2.346\\
$^{65}{\rm Cu}$ &$p_{3/2}$ &  2.3816 (2)& 3.793 &2.398   &  2.759 &0.23&0.24 &  2.460\\
$^{67}{\rm Cu^{c)ÿ}}$ &$p_{3/2}$ & 2.54(2)& 3.793  &  ---    &  2.781 &0.14&0.21 &  2.536\\
$^{69}{\rm Cu}$ &$p_{3/2}$ & 2.84(2)& 3.793  &  ---    &  2.789 &0   &0.075&  2.789\\
$^{71}{\rm Cu}$ &$p_{3/2}$ & 2.28(1)& 3.793   & ---    &  2.778 &0.30&0.26 &  2.444\\
$^{73}{\rm Cu}$ &$p_{3/2}$ & \cite{UliKos}& 3.793  & ---    &  2.752 &$\simeq$0.60&--- &  ---\\
\end{tabular}
\end{ruledtabular}
\end{table}

\subsection{Calculations for copper chain assuming deformation}
Let us now carry out an alternative calculation supposing these Cu
isotopes to be deformed. To estimate the possible impact of the
quadrupole deformation on the magnetic moments one may assume a
simple rigid rotor picture \cite{Bohr} with the effective
rotational g-factor being the one for a uniformly charged rigid
body $g_R=Z/A$ and the intrinsic g-factor $g_K=\mu_j/j$, where
$\mu_j$ stands for the magnetic moment of the odd quasiparticle in
the spherical core. In fact, such an approximation accounts only
for a change of the kinematic factors in the case of a deformed
core due to the precession of the rotational and intrinsic angular
momenta with respect to the total angular momentum. As seen from
Fig.~\ref{Cu} (open circles labelled with $|3/1^{-}, K=3/2\rangle$
(deform), this simplest model leads to a qualitative agreement
with the data for the nuclei at the left wing of the chain, i.e.
for the neutron-deficient $^{57,59}$Cu isotopes. As to the right
wing with the neutron-rich $^{73-77}$Cu isotopes, the situation is
less clear. Only for $^{73}$Cu ($I^{\pi}=3/2^-$) an experimental
moment value is available and the ``deformed'' calculation
practically agrees with the experiment. Note that the systematic
calculations \cite{HF} predict a significant $\beta_2$ deformation
for  the $^{73-77}$Cu isotopes. At the same time, according to
this systematic, the nuclei in the middle of the shell ($63<A<73$)
are relatively weakly deformed. Therefore, this rough model is
obviously not justified for those nuclei. On the other hand, the
fact that the many-particle SM calculations for the three Cu
isotopes in the vicinity of N=40  (i.e. A = 61-65; filled circles
in Fig.~\ref{Cu}) \cite{Brown} are in agreement with the
experimental data may give evidence that these nuclei are nearly
spherical  but that the weight of many-particle configurations in
the ground-state is rather high. It were very instructive to carry
out similar many-particle SM calculations for the other Cu
isotopes belonging to  the $2p-1f$ shell.

\begin{table}
 \caption{Main 2$_1^+$-state characteristics in even Ni isotopes
  \cite{BE2,Sor,Per}\label{Ni}}

\begin{ruledtabular}
\begin{tabular}{cccc}
 A & $\omega_{2_1^+}$ (keV)
 & B(E2)$\uparrow$ (W.U.)&$\beta$\\
\hline 56& 2700.6(7) & 9.4(19)&
0.173(17) \\  58& 1454.0(1)&  10.4(3)&  0.1828(26)\\
 60& 1332.518(5)&  13.4(2)&  0.2070(17)\\
 62& 1172.91(9)& 12.2(3)&  0.1978(28)\\
 64& 1345.75(5)&  10.0(11)&  0.179(9) \\
66& 1425.1(3)&  7.8(11)&  0.158(12)\\
68& 2033.2(2)&  3.2(7)&  0.100(12)\\
 70& 1259.6(2)&9.9(1.6)&0.178(15)\\
\end{tabular}
\end{ruledtabular}
\end{table}

\subsection{The nuclei at the extremes of the Cu chain}

A different situation takes place in the vicinity of the even-even
core $^{56}$Ni which could in principle be considered as ``doubly
magic'',
but in fact $N=28$ and  $Z=28$ turn out to be a bad sequence of
magic numbers.   Spectroscopic studies \cite{Rud,Johan}  have
revealed evidence for two well-deformed excited bands in
$^{56}$Ni. The first rotational band can be explained within
shell-model calculations in the full pf model space while the
second one seems to be related to the excitations into the
$1g_{9/2}$-orbit \cite{Horoi}. Actually, the isotope $^{56}$Ni is
so ``soft'' that it becomes deformed by addition of only one
proton. As a result, the rotor-particle model explains the
experimental value of its magnetic moment. In isotopes with  
large neutron excess, i.e. $A\geq$73, the deformation pattern is
not so definite. Absence of the experimental data for $\omega_2$
and $B(E2)$ in the corresponding Ni isotopes does not permit to
find the magnetic moment within  the Bohr-Mottelson model. An
attempt to use the phonon-particle model in this mass region shows
that the adjusted value of the phonon admixture coefficient turns
out to be unrealistically large (see Table~\ref{Cu}). On the other
hand, the particle-rotor model also explains the data only
qualitatively. For more definite conclusions, it were instructive
to examine the properties of the 2$^+$-excitation in the  core,
 $^{76}$Ni, which is very close to the 3-rd semi-magic isotope
 $^{78}$Ni. In addition, it were important to determine the
 $I^{\pi}$ values  as well as the experimental
 magnetic moments for the $^{75,77}$Cu nuclei.

Thus, with the above reservations, we see that examination of the
magnetic moments of the long Cu isotope chain permits to retrace
the evolution of the  shape of the ground state, from spherical in
the middle of the chain to deformed at the wings where the nuclei
have a large neutron excess or neutron deficiency.

\section{Conclusions and discussion}
We have performed a systematic analysis of  the magnetic moments
of several long isotope and isotone chains of medium heavy and
heavy nuclei. The modern version of the self-consistent FFS theory
with exact account for the continuum has been used. The
quasi-particle self-consistent basis is generated within the
Generalized EDF method \cite{Fay}. The extended Landau-Migdal
effective interaction includes the spin-isospin tensor terms
induced  by the one-pion and one-rho-meson exchange. For the vast
majority of the nuclei considered  in the present work and in
\cite{BST}  good agreement at the level of 0.1-0.2$\mu_N$ has been
obtained. The best accuracy was achieved for the chains of the
semi-magic lead and tin isotopes.

We analyzed also in detail the isotopic copper chain for which new
data on magnetic moments  were obtained recently
\cite{Isolde_Cu,Cu_1}. It includes  nuclei far from the
$\beta$--stability valley, with a maximal relative neutron excess,
$\eps_+ = 33\%$. Our study  has shown that the analysis of
magnetic moments of long isotopic chains of non-magic nuclei
provides important information on the evolution of the ground
state structure with neutron number.
  Copper nuclei can be described in terms of a proton
added to the even-even Ni core. If, with some reservations, one
considers $N=28$ and $Z=28$ as magic numbers, the Ni core chain
will include two magic nuclei, $^{56}$Ni and $^{68}$Ni,  ending
with the isotope $^{76}$Ni which is very close to the next magic
nucleus $^{78}$Ni. In the middle of the chain, in the vicinity of
the $^{68}$Ni core, magnetic moments are sufficiently well
reproduced supposing the spherical symmetry of the ground state if
 the $2^+_1$-state virtual admixture  is taken into account. In the
wings of the chain predictions of the theory with the spherical
basis differ  significantly from the data. Qualitative agreement
could be achieved when supposing the ground state to be deformed,
in accordance with systematic calculations \cite{HF}. Thus, a
sharp change of the magnetic moment value under addition of a few
nucleons may give a signal of a sudden change in nuclear
structure. It should be noted that the situation for the
$^{75,77}$Cu isotopes is not clear as far as the ground state
$I^{\pi}$ values are not definitely known experimentally in this
case.

It is worth to stress that our conclusions about the structure of
nuclei of the Cu chain has to be considered as preliminary in view
of the fact that rather schematic models were used for the surface
vibration admixture in the spherical case and also for deformed
nuclei. It would be instructive to carry out calculations for the
seven members of the chain which belong to the $2p-1f$ shell
within the approach of \cite{Brown}. It would be also of interest
to perform the calculation of the magnetic moments within the
hybrid approach uniting the many-particle SM and the FFS theory.
So far, this method \cite{SKh} has been applied only for the
isolated 1$f_{7/2}$ shell.

 It is interesting also to mention the possibility of a new type of phase
transitions in nuclei which has recently been predicted by V.A.
Khodel et al. \cite{KhC}.  It consists in a rearrangement of the
Fermi system vacuum due to  the merging of a pair of
single-particle levels close to the Fermi surface.  This phase
transition results in partial occupation numbers and is analogous
to the so-called fermionic condensation in infinite Fermi systems
\cite{KhSh}. According to estimations in \cite{KhC}, such phase
transition could occur in the $2p-1f$ nuclei.  It would be of
great interest to carry out an analysis of the magnetic moments
supposing that such phase transition takes place.

The authors thank V.A. Khodel, A.N. Andreev and O.I. Ivanov and
Y.A. Litvinov for valuable discussions and U. Koster for sending
us his unpublished data.

This research was partially supported by the joint Grant of the
Russian Fund for Basic Research (RFBR) and Fund for Scientific
Research Flanders RFBR-Fl-05-02-19813, by the Grant
NSh-3004.2008.2  of the Russian Ministry for Science and
Education, by the RFBR grants 06-02-17171-a, 07-02-00553-a and by
DFG, Germany via the contract No. 436 RUS 113907/0-1.

{}

\begin{thebibliography}{99}


\bibitem{Bohr} A.~Bohr and B.~R.~Mottelson, {\it Nuclear Structure} (Benjamin,
New York, 1971.), Vol. 1

\bibitem{AB1} A.~B.~Migdal, {\it Theory of finite Fermi systems and applications to
atomic nuclei} (Wiley, New York, 1967).

\bibitem{Stone} N.~J.~Stone,  ADNDT, {\bf 90}, 75 (2005).

\bibitem{BST}  I.~N.~Borzov, E.~E.~Saperstein, S.~V.~Tolokonnikov,
Phys. At. Nucl., to be published.

\bibitem{STF}
A.~V.~Smirnov, S.~V.~Tolokonnikov, S.~A.~ Fayans, Sov. J. Nucl.
Phys. {\bf 48}, 1030 (1988).

\bibitem{Fay}
S.~A.~Fayans, S.~V.~Tolokonnikov, E.~L.~Trykov, and D.~Zawischa,
Nucl.~Phys. {\bf A676}, 49 (2000).

\bibitem{ShB} S.~Shlomo, G.F.~Bertsch, Nucl. Phys., {\bf A243},
507 (1975).

\bibitem{SapTF}  E.~E.~Saperstein, S.~V.~Tolokonnikov, S.~A.~Fayans,
Preprint KIAE-2571 (1975).

\bibitem{Plat} A.~P.~Platonov, E.~E.~Saperstein, Nucl. Phys. {\bf A486}, 63 (1988).

\bibitem{PF}
N.~I.~Pyatov and S.~A.~Fayans. Sov. J. Part. Nucl. {\bf 14}, 401
(1983).

\bibitem{BorTF}  I.~N.~Borzov, S.~V.~Tolokonnikov, and S.~A.~Fayans.
    Sov. J. Nucl. Phys. {\bf 40},  732 (1984).


\bibitem{Minam}
K. Minamisono et al., Phys.Rev.Lett. {\bf 96}, 102501 (2006)

\bibitem{Isolde_Cu}
 V.~V.~ Golovko, I.~Kraev, T.~Phalet, N.~Severijns et al.,
Phys. Rev. {\bf C70}, 014312 (2004).


\bibitem{Cu_1} K.~T.~Flanagan, G.~Neyens et al. (in preparation).

\bibitem{UliKos} U.~Koster et al. (p.c., unpublished ISOLDE-RILIS data).

\bibitem{Pb} M.~D.~Seliverstov et al. (in preparation).

\bibitem{KS} W.~Kohn, L.~J.~Sham, Phys. Rev. {\bf A140}, 1133 (1965).


\bibitem{HF} Y.~Aboussir, N.~Pearson, A.~K.~Dutta, and F.~Tondeur, ANDT {\bf 61}, 127 (1995).

\bibitem{Dutta} S.~B.~Dutta, R.~Kirchner, O.~Klepper et al., Z. Phys. A {\bf 341}, 39 (1991).

\bibitem{Dinger} U.~Dinger, J.~Eberz, G.~Huber et al., Z. Phys. A {\bf 328}, 253 (1987).

\bibitem{Anselment} M.~Anselment, W.~Faubel, S.~Goring et al.
Nucl. Phys. {\bf A451}, 471 (1986).

\bibitem{Bo}  I.~N.~Borzov, Nucl. Phys. {\bf A777}, 435 (2006).

\bibitem{Go12} B.~I.~Gorbachev, A.~I.~Levon, O.~F.~Nemts et al.,
Zh. Eksp. Teor. Fiz. {\bf 87}, 3 (1984).

\bibitem{Ej01} H.~Ejiri, T.~Shibata, M.~Takeda, Nucl. Phys. {\bf
A221} 211 (1974).

\bibitem{Pr02} H.~Prade, L.~Kaubler, U.~Hageman et al., Nucl.
Phys. {\bf A333}, 33 (1980).

\bibitem{Brown}
M.~Honma, T.~Otsuka, B.~A.~Brown, T.~Mizusaki, Phys. Rev. {\bf
C69}, 034335 (2004)

\bibitem{Kaneko}
K.~Kaneko, M.~Hasegawa, T.~Mizusaki, Y.~Sun, Phys. Rev. {\bf C74},
024321 (2006)

\bibitem{KhS} V.~A.~Khodel and E.~E.~Saperstein, Phys. Rep. {\bf
92}, 183 (1982).

\bibitem{Bohr2} A.~Bohr and B.R.~Mottelson, {\it Nuclear Structure} (Benjamin,
New York, Amsterdam, 1974.), Vol. 2.

\bibitem{Hammam}
I.~Hammamoto, Phys. Lett. {\bf B61}, 343 (1973).

\bibitem{RiStone}
J. Rikovska and N.J. Stone, Hyp. Int. {\bf129}, 131 (2000).

\bibitem{Rud}
D.~Rudolph, C.~Baktash, M.J.~Brinkman, E. Caurier  et al., Phys.
Rev. Lett. {\bf 82}, 3763 (1999).

\bibitem{Johan}
E.K. Johansson et al., Eur.Phys.J. {\bf A 27}, 157 (2006).

\bibitem{Horoi}
M.~Horoi, B.~Brown, T.~Otsuka et al., Phys. Rev. {\bf C73}, 061305
(2006).

\bibitem{BE2}
NNDC Databases, http://www.nndc.bnl.gov/be2

\bibitem{Sor}
O.~Sorlin, S.~Leenhardt, C.~Donzaud et al., Phys. Rev. Lett. {\bf
88}, 092501 (2002).

\bibitem{Per}
O.~Perry, O.~Sorlin, S.~Franchoo et al., Phys. Rev. Lett. {\bf
96}, 232501 (2006).


\bibitem{SKh} E.E.~Saperstein, V.A.~Khodel, Yad. Fiz., {\bf 4}, 701
(1966).


\bibitem{KhC} V.~A.~Khodel, J.~W.~Clark, Haotchen Li, and M.~V.~Zverev,
Phys. Rev. Lett. {\bf 98}, 216404 (2007).

\bibitem{KhSh} V.~A.~Khodel, V.~R.~Shaginyan,
JETP Lett.,{\bf 51}, 553 (1990).



\end{thebibliography}
\end{document}